\documentclass[aps,prl,reprint,superscriptaddress,longbibliography]{revtex4-2}
\usepackage[colorlinks=true,
            linkcolor=blue,
            citecolor=blue,
            urlcolor=blue]{hyperref}
\usepackage[T1]{fontenc}
\usepackage[utf8]{inputenc}
\usepackage{lmodern}
\usepackage{amsmath,amssymb,bm,mathtools}
\usepackage{graphicx}
\usepackage{xcolor}
\usepackage{booktabs}
\usepackage{microtype}
\usepackage{hyperref}
\usepackage{tikz}
\usetikzlibrary{arrows.meta,positioning,calc,decorations.pathmorphing}
\hypersetup{colorlinks=true,linkcolor=blue,citecolor=blue,urlcolor=blue}
\usepackage{xcolor}
\usepackage[normalem]{ulem}

\newcommand{\vk}{\mathbf{k}}

\begin{document}

\title{Two-Color Coherent Control of the Spin Photovoltaic Effect in Centrosymmetric Crystals}

\author{Songmin Liu}
\thanks{These authors contributed equally to this work.}
\affiliation{Key Laboratory of Low-dimensional Materials and Application Technology, 
School of Materials Science and Engineering, Xiangtan University, 
Xiangtan 411105, China}

\author{Pan Zhou$^{*}$}
%\thanks{These authors contributed equally to this work.}
\email{zhoupan71234@xtu.edu.cn}
\affiliation{Key Laboratory of Low-dimensional Materials and Application Technology, 
School of Materials Science and Engineering, Xiangtan University, 
Xiangtan 411105, China}

\author{Lizhong Sun}
\email{lzsun@xtu.edu.cn}
\affiliation{Key Laboratory of Low-dimensional Materials and Application Technology, 
School of Materials Science and Engineering, Xiangtan University, 
Xiangtan 411105, China}
\affiliation{Hunan Provincial Key laboratory of Thin Film Materials and Devices, 
School of Materials Science and Engineering, Xiangtan University, 
Xiangtan 411105, China}

\begin{abstract}
Spatial inversion forbids the conventional electric-dipole second-order bulk charge and spin photovoltaic responses in centrosymmetric crystals. Here we show that a phase-coherent $\omega$--$2\omega$ two-color field can generate symmetry-selected charge and spin photocurrents through an inversion-even third-order dc response. More importantly, the polarization direction and relative phase of the two-color field provide direct control over the current direction and can isolate transport channels in which the charge current vanishes while a finite spin current remains. A systematic symmetry analysis of centrosymmetric magnetic point groups and collinear antiferromagnetic spin point groups shows that pure-spin channels are allowed in a broad class of such systems under appropriate two-color driving. Furthermore, the relative optical phase governs both the magnitude and sign of the response. Our calculations on inversion-stacked bilayer $2H$-MoS$_2$ and a centrosymmetric collinear altermagnet demonstrate resonantly enhanced, phase-tunable spin photocurrents in these complementary systems with strong and weak spin--orbit coupling, respectively. Our results establish the polarization direction and phase of coherent two-color light as active control parameters for spin transport in centrosymmetric materials.
\end{abstract}

\maketitle
\begin{figure}[t]
    \centering
    \includegraphics[width=\linewidth]{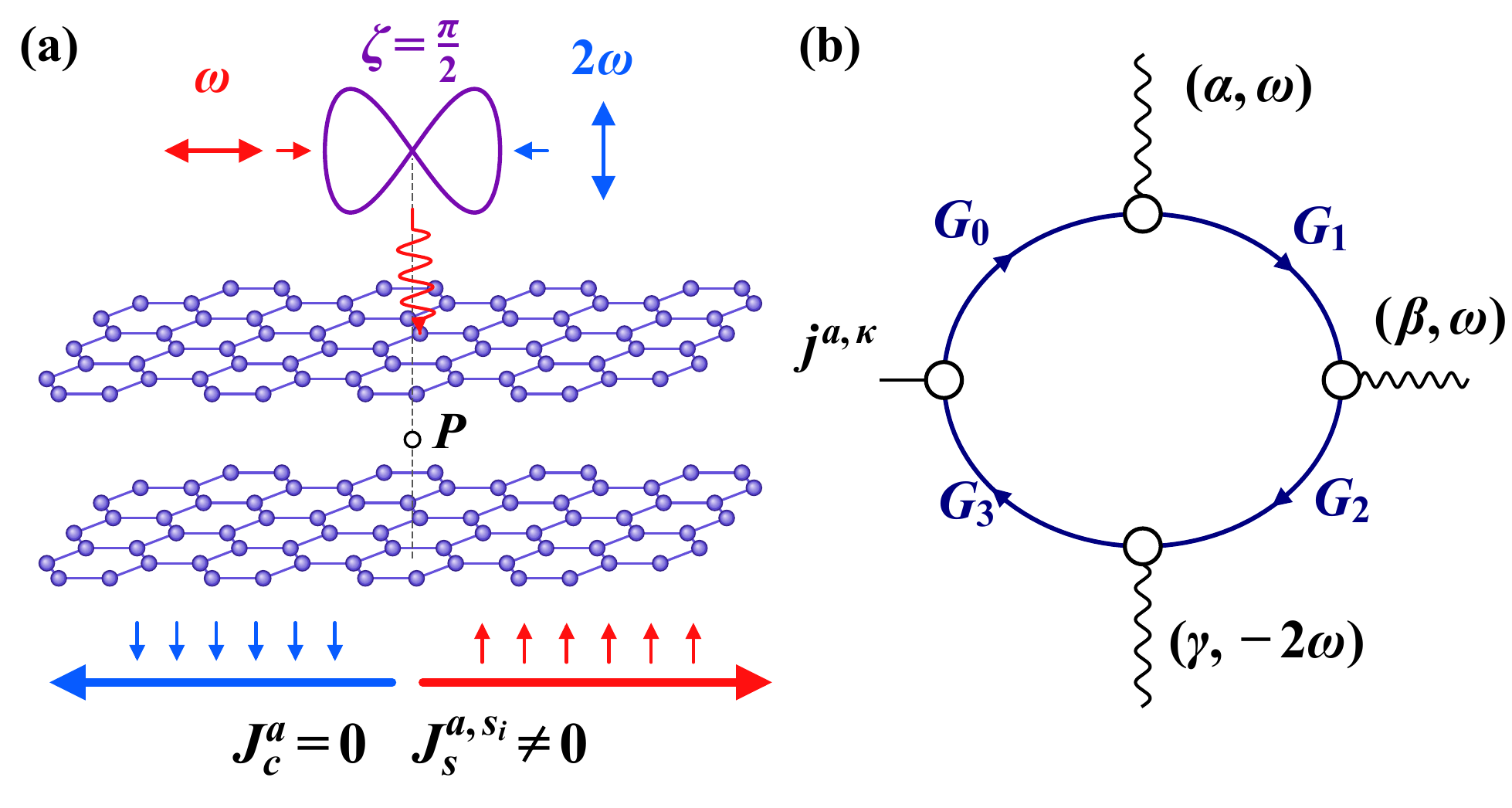}
    \caption{
    (a) Schematic of a pure spin photocurrent induced by phase-locked,
    orthogonally linearly polarized fields at $\omega$ and $2\omega$
    in a centrosymmetric crystal.
    Counterpropagating carriers with opposite spin polarizations
    cancel the charge current along the selected direction,
    $J_c^a=0$, while producing a finite spin current,
    $J_s^{a,s_i}\neq0$.
    The inversion center is denoted by $P$.
    (b) Representative three-linear-vertex Keldysh diagram for the
    cubic dc channel $\omega+\omega-2\omega=0$.
    The optical vertices are labeled by $(\alpha,\omega)$,
    $(\beta,\omega)$, and $(\gamma,-2\omega)$, and the measured-current
    vertex is $j^{a,\kappa}$.
    The directed propagator segments are labeled $G_0$--$G_3$.
    }
    \label{fig:lpl_mechanism}
\end{figure}

%%%%%%%%%%%%%%%%%%%%%%%%%%%%%%%%%%%%%%%%%%%%%%%%%%%%%%%%%%%%%%%%%%
\indent\textit{Introduction.---} The generation and control of spin currents are central objectives of spintronics~\cite{vzutic2004,Sinova2015,Baltz2018}, with optical excitation providing an ultrafast, contact-free route to manipulating spin transport~\cite{Hubner2003,Bhat2005,Zhao2005,Kirilyuk2010,Nemec2018,YangLiang2026}. A prominent example of optically driven transport is the second-order bulk photovoltaic effect (BPVE), which generates dc charge currents in noncentrosymmetric crystals~\cite{Kraut1979,Belinicher1980,vonBaltz1981,Aversa1995,Sipe2000,Fridkin2001,Young2012,Rangel2017,Morimoto2016,ahn2020low,Dai2023,Ma2021,Ahn2022,Morimoto2023}. Beyond charge transport, second-order optical responses can also generate spin photocurrents and, under appropriate symmetry conditions, pure spin currents without accompanying charge flow~\cite{Xu2021,Hsu2025}. These phenomena have been extensively explored in magnetic, topological, and low-dimensional materials through mechanisms involving spin-dependent shift and injection currents, as well as circular photogalvanic responses~\cite{Young2013,Xiao2021,Fei2021,Xiao2023,Dong2025,Yoshida2026,Zhou2007,Kim2017,Huang2017,Hamamoto2017,Ishizuka2019PRL,Ishizuka2019PRB,Song2021}. However, these conventional second-order responses face a fundamental symmetry constraint: within the electric-dipole approximation, spatial inversion reverses both charge and spin currents while leaving the quadratic electric-field products unchanged. Consequently, homogeneous second-order charge and spin photocurrents are forbidden in centrosymmetric crystals.\\
\indent Third-order nonlinear responses offer a natural route to overcoming this symmetry constraint~\cite{Atanasov1996,Bhat2000,Stevens2003,Ezawa2026,Zhang2026ThirdOrder}: because charge and spin currents, like cubic products of the electric field, are odd under spatial inversion, third-order dc photocurrents are allowed even in centrosymmetric crystals. For charge transport, bicircularly polarized light generates phase-dependent photocurrents through quantum interference between distinct excitation pathways~\cite{Ikeda2023,Guo2025}. For spin transport, monochromatic optical excitation combined with an external dc electric field has been shown to produce pure spin currents in centrosymmetric altermagnets through the interplay of band geometry and crystal symmetry~\cite{Dong2026}. These mechanisms, however, either yield charge rather than spin currents or rely on an external dc bias. It thus remains an open question whether polarized light alone can provide a general, symmetry-based mechanism for pure spin currents in centrosymmetric materials---with or without spin-orbit coupling (SOC)---while symmetry forbids any accompanying charge current.\\
\indent In this work, we show that phase-coherent $\omega$--$2\omega$ two-color fields (TCFs) provide an all-optical route to pure spin photocurrents in centrosymmetric systems. We derive the corresponding third-order charge and spin photoconductivities and establish their symmetry constraints for both spin--orbit-coupled systems described by magnetic point groups and nonrelativistic collinear magnets described by spin point groups. For a broad class of centrosymmetric symmetries, an appropriate orthogonal TCFs suppresses the charge current along a selected transport direction while allowing a finite spin current. The relative optical phase further provides a continuous control of the magnitude and sign of the response. We illustrate these principles in inversion-stacked bilayer $2H$-MoS$_2$ and a centrosymmetric collinear altermagnet, representing complementary relativistic and nonrelativistic regimes. Our results establish coherent two-color light as a general means of generating and controlling pure spin photocurrents in centrosymmetric quantum materials.\\
%%%%%%%%%%%%%%%%%%%%%%%%%%%%%%%%%%%%%%%%%%%%%%%%%%%%%%%%%%%%%%%%%%
\indent\textit{Two-color field and induced spin photocurrent.---} We consider a phase-locked two-color optical field composed of linearly polarized components at frequencies $\omega$ and $2\omega$. Using the Fourier convention $\mathbf E(t)=\sum_\nu\mathbf E(\nu)e^{\mathrm{i}\nu t}$, we write
\begin{align}
\mathbf E(\omega)
&=
E_\omega e^{\mathrm{i}\phi_\omega}
\mathbf u,
&
\mathbf E(-2\omega)
&=
E_{2\omega}e^{-\mathrm{i}\phi_{2\omega}}
\mathbf v,
\label{eq:two_color_field}
\end{align}
with $\mathbf E(-\nu)=\mathbf E(\nu)^*$. Here $E_\omega$ and $E_{2\omega}$ are nonnegative Fourier amplitudes, $\mathbf u = \hat{\mathbf e}_\omega$ and $\mathbf v = \hat{\mathbf e}_{2\omega}$ are real unit polarization vectors, and $\phi_\omega$ and $\phi_{2\omega}$ are the phases of the positive-frequency components. Although the two polarization directions can be varied independently, we focus on orthogonal fields, $\mathbf u\cdot\mathbf v=0$.\\
\indent In the spatially uniform electric-dipole limit, inversion symmetry forbids the conventional second-order dc charge and spin photovoltaic responses~\cite{Sipe2000,Xu2021}. The leading symmetry-allowed dc response is therefore cubic, arising from the frequency combination $\omega+\omega-2\omega=0$. The relative phase relevant to the dc response is $\zeta=2\phi_\omega-\phi_{2\omega}$, which is independent of the choice of time origin and it is a tunable degree of freedom for two-color light. Our aim is to identify polarization configurations that produce a finite spin current without an accompanying charge current along a selected transport direction, $J_c^a=0$ and $J_s^{a,s_i}\neq0$, as illustrated in Fig.~\ref{fig:lpl_mechanism}(a).\\
\indent We describe the charge- and spin-current responses in terms of the corresponding current vertices,
\begin{equation}
j^{a,0}=v^a,
\qquad
j^{a,s_i}=\frac{1}{2}\{v^a,s_i\},
\qquad
s_i=\frac{\hbar}{2}\sigma_i,
\label{eq:lpl_current_vertices}
\end{equation}
where the superscripts $0$ and $s_i$ label the charge and spin channels, respectively~\cite{Sinova2004,Shi2006,Yoshida2026}. For a spatially uniform optical field, the vector potential enters through the Peierls substitution,
\begin{equation}
H_{\mathbf A}(\mathbf k)
=
H_0\!\left(
\mathbf k+\frac{e\mathbf A}{\hbar}
\right),
\qquad
A_\lambda(\nu)
=
\frac{\mathrm{i}}{\nu}E_\lambda(\nu).
\label{eq:lpl_peierls}
\end{equation}
We evaluate the third-order response in the Keldysh formalism, retaining the field dependence of both the Hamiltonian and the measured-current operator~\cite{Passos2018,Parker2019}.\\
\indent To illustrate its microscopic structure, we focus on the primary contribution arising from the three-linear-vertex 0;111 sector. Temporarily labeling the optical legs as
\begin{equation}
1=(\alpha,\omega),
\qquad
2=(\beta,\omega),
\qquad
3=(\gamma,-2\omega),
\label{eq:lpl_optical_legs}
\end{equation}
the ordering $1\rightarrow2\rightarrow3$ gives
\begin{widetext}
\begin{align}
\sigma_{0;1|2|3}^{a,\kappa;\alpha\beta\gamma}
={}&
\frac{\mathrm{i}e^3}{2\omega^3}
\int_{\mathrm{BZ}}
\frac{\mathrm{d}^d k}{(2\pi)^d}
\sum_{nmlp}
\frac{j_{mn}^{a,\kappa}}{\mathcal D_{nm}(0)}
\Bigg\{
\frac{v_{nl}^{\gamma}}{\mathcal D_{lm}(2\omega)}
\left[
\frac{f_{pm}v_{lp}^{\beta}v_{pm}^{\alpha}}
{\mathcal D_{pm}(\omega)}
-
\frac{f_{lp}v_{lp}^{\alpha}v_{pm}^{\beta}}
{\mathcal D_{lp}(\omega)}
\right]
\notag\\
&\hspace{2.2cm}
-
\frac{1}{\mathcal D_{nl}(2\omega)}
\left[
\frac{f_{pl}v_{np}^{\beta}v_{pl}^{\alpha}}
{\mathcal D_{pl}(\omega)}
-
\frac{f_{np}v_{np}^{\alpha}v_{pl}^{\beta}}
{\mathcal D_{np}(\omega)}
\right]
v_{lm}^{\gamma}
\Bigg\},
\label{eq:lpl_sigma_0111}
\end{align}
\end{widetext}
where
\begin{equation}
\mathcal D_{ab}(r\omega)
=
\varepsilon_b-\varepsilon_a-r\hbar\omega+\mathrm{i}\eta,
\qquad
f_{ab}=f_a-f_b ,
\label{eq:lpl_denominator}
\end{equation}
and $\eta>0$ specifies the relaxation broadening. All band energies and matrix elements are evaluated at the same crystal momentum. The remaining linear-vertex orderings, the higher-order input contact vertices, and the field-dependent current-vertex terms are all retained in the numerical calculations. These contributions must be combined to eliminate spurious low-frequency behavior caused by an incomplete vertex expansion~\cite{Aversa1995,Passos2018,Parker2019}. Their explicit expressions and counting conventions are given in the Supplemental Material (SM).\\
\indent We denote the complete coefficient multiplying $E_\alpha(\omega)E_\beta(\omega)E_\gamma(-2\omega)$ by $S_{\mathrm{TCF}}^{a,\kappa;\alpha\beta\gamma}$. Because the two fundamental-frequency fields are identical, only the part symmetric in their optical indices contributes,
\begin{equation}
S_{\mathrm{TCF}}^{a,\kappa;(\alpha\beta)\gamma}
=
\frac{1}{2}
\left[
S_{\mathrm{TCF}}^{a,\kappa;\alpha\beta\gamma}
+
S_{\mathrm{TCF}}^{a,\kappa;\beta\alpha\gamma}
\right].
\label{eq:lpl_intrinsic_permutation}
\end{equation}
For the linearly polarized fields in
Eq.~\eqref{eq:two_color_field}, the relevant coefficient is the polarization projection
\begin{equation}
\mathcal S^{a,\kappa}
=
S_{\mathrm{TCF}}^{a,\kappa;(\alpha\beta)\gamma}
u_\alpha
u_\beta
v_\gamma .
\label{eq:lpl_helicity_projection}
\end{equation}
Unlike a circular-polarization projection, all polarization-vector components in this expression are real; the optical phase dependence is carried entirely by $e^{\mathrm{i}\zeta}$.\\
\begin{table}[t]
\squeezetable
\caption{
Symmetry transformations of the third-order
 numerators $\mathcal N_{0}$ and $\mathcal N_{s_i}$.
Here $N_x$ is the number of $x$ indices among
$(a,\alpha,\beta,\gamma)$. In the first column, the first four operations correspond to magnetic point group operations, while the last four correspond to spin group operations.
}
\label{tab:third_order_symmetry}
\begin{ruledtabular}
\begin{tabular}{lcc}
Operation
&
Charge kernel
&
Spin kernel
\\
\hline

$\mathcal P$
&
$+\mathcal N_0(-\vk)$
&
$+\mathcal N_{s_i}(-\vk)$
\\

$\mathcal T$
&
$+\mathcal N_0^{*}(-\vk)$
&
$-\mathcal N_{s_i}^{*}(-\vk)$
\\

$\mathcal M_x$
&
$(-1)^{N_x}\mathcal N_0(\mathcal M_x\vk)$
&
$\begin{array}{c}
(-1)^{N_x}\mathcal N_{s_x}(\mathcal M_x\vk)
\\[-1mm]
-(-1)^{N_x}\mathcal N_{s_{y,z}}(\mathcal M_x\vk)
\end{array}$
\\

$\mathcal M_x\mathcal T$
&
$(-1)^{N_x}\mathcal N_0^{*}(-\mathcal M_x\vk)$
&
$\begin{array}{c}
-(-1)^{N_x}\mathcal N_{s_x}^{*}(-\mathcal M_x\vk)
\\[-1mm]
+(-1)^{N_x}\mathcal N_{s_{y,z}}^{*}(-\mathcal M_x\vk)
\end{array}$
\\

$[\bar C_2\Vert\mathcal T]$
&
$+\mathcal N_0^{*}(-\vk)$
&
$+\mathcal N_{s_z}^{*}(-\vk)$
\\

$[C_2\Vert\mathcal P]$
&
$+\mathcal N_0(-\vk)$
&
$-\mathcal N_{s_z}(-\vk)$
\\

$[E\Vert\mathcal M_x]$
&
$(-1)^{N_x}\mathcal N_0(\mathcal M_x\vk)$
&
$(-1)^{N_x}\mathcal N_{s_z}(\mathcal M_x\vk)$
\\

$[C_2\Vert\mathcal M_x]$
&
$(-1)^{N_x}\mathcal N_0(\mathcal M_x\vk)$
&
$-(-1)^{N_x}\mathcal N_{s_z}(\mathcal M_x\vk)$
\\

\end{tabular}
\end{ruledtabular}
\end{table}
\indent\textit{Symmetry conditions for pure spin photocurrents.---} We now establish the symmetry conditions under which the third-order TCF response can carry a finite spin current while the accompanying charge current vanishes. For the symmetry analysis, it is sufficient to track the transformation of the matrix-element numerator,
\begin{equation}
\mathcal N_{\kappa}^{a\alpha\beta\gamma}(\vk)
=
(j^{a,\kappa})_{mn}
v^\gamma_{nl}
v^\beta_{lp}
v^\alpha_{pm},
\qquad
\kappa=0,s_i ,
\label{eq:numerator}
\end{equation}
where \(j^{a,0}=v^a\) and \(j^{a,s_i}=\{v^a,s_i\}/2\). The energy denominators and occupation factors are scalars under the unitary spatial operations considered below, while antiunitary operations additionally complex conjugate the corresponding matrix elements. The transformation properties of the relevant numerators are summarized in Tab.~I.\\
\indent We first consider inversion-symmetric systems with SOC, for which spin transforms as an axial vector under crystalline operations. Under spatial inversion, \(v^a(\vk)\rightarrow-v^a(-\vk)\), whereas \(s_i(\vk)\rightarrow+s_i(-\vk)\).  Since both the charge-current and spin-current vertices are inversion odd, the three additional optical velocity vertices give
\begin{equation}
\mathcal P:\qquad
\mathcal N_{\kappa}^{a\alpha\beta\gamma}(\vk)
\rightarrow
+\mathcal N_{\kappa}^{a\alpha\beta\gamma}(-\vk).
\label{eq:P_parity}
\end{equation}
Thus, in contrast to the conventional second-order BPVE and bulk spin photovoltaic effects, the third-order TCF charge and spin kernels are both inversion even, and inversion symmetry does not forbid the response. However, time reversal acts differently on charge and spin currents,
\begin{align}
\mathcal T:\qquad
\mathcal N_0(\vk)
&\rightarrow
+\mathcal N_0^{*}(-\vk),
\nonumber\\
\mathcal N_{s_i}(\vk)
&\rightarrow
-\mathcal N_{s_i}^{*}(-\vk).
\label{eq:T_parity}
\end{align}
These antiunitary relations place charge and spin kernels in opposite symmetry sectors. However, time-reversal symmetry alone is insufficient to generate a spin current in an inversion-symmetric system. To further constrain the system, we consider the presence of a mirror operation $\mathcal M_x$, under which the velocity and spin vectors transform as
\begin{align}
(v^x,v^y,v^z)&\rightarrow(-v^x,v^y,v^z),
\nonumber\\
(s_x,s_y,s_z)&\rightarrow(s_x,-s_y,-s_z).
\label{eq:Mx_vectors}
\end{align}
Let \(N_x\) denote the number of \(x\) indices among
\((a,\alpha,\beta,\gamma)\).  The corresponding numerator transforms as
\begin{align}
\mathcal N_0(\vk)
&\rightarrow
(-1)^{N_x}
\mathcal N_0(\mathcal M_x\vk),
\nonumber\\
\mathcal N_{s_x}(\vk)
&\rightarrow
(-1)^{N_x}
\mathcal N_{s_x}(\mathcal M_x\vk),
\nonumber\\
\mathcal N_{s_{y,z}}(\vk)
&\rightarrow
-(-1)^{N_x}
\mathcal N_{s_{y,z}}(\mathcal M_x\vk).
\label{eq:Mx_parity}
\end{align}
For odd \(N_x\), the charge numerator is mirror odd, whereas the spin-\(y\) and spin-\(z\) numerators are mirror even. Integration over mirror-related momenta therefore gives
\begin{equation}
S^{a,0}=0,
\qquad
S^{a,s_{y,z}}
\ \text{allowed},
\label{eq:pure_spin_SOC}
\end{equation}
providing a symmetry-selected pure-spin tensor channel. For even \(N_x\), the complementary rule holds: the spin-\(y,z\) components are forbidden, whereas charge and spin-\(x\) components are symmetry allowed. By analyzing all 32 magnetic point groups that contain spatial inversion, we find that 25 of them include a mirror plane along some direction (see the SM for details). This shows that materials belonging to these magnetic point groups can generate a pure spin current under illumination by TCF along mirror-related specific directions. Of course, it must be emphasized that for a given point group the definition of the coordinate axes—in particular the orientation of the mirror plane—may differ, so that the polarization directions of the light fields must be adjusted accordingly. Moreover, besides the point group that possesses only spatial inversion, every other inversion-containing point group admits suitable light-field polarization directions that can induce a spin current, and the detailed results are given in SM.\\
\indent We next consider the nonrelativistic collinear limit of altermagnets, where spin and spatial transformations are independent and the appropriate constraints are described by spin-group operations~\cite{Smejkal2022,Liu2022SpinGroup,Sivianes2025,Schiff2025,alter1,alter2,alter3,alter4,alter5}. Taking the collinear axis along \(z\), \(s_z\) is conserved and the TCF response can be decomposed into the two spin channels,
\begin{align}
J^{a,0}_{\rm TCF}
&=
J^{a,\uparrow}_{\rm TCF}
+
J^{a,\downarrow}_{\rm TCF},
\nonumber\\
J^{a,s_z}_{\rm TCF}
&=
J^{a,\uparrow}_{\rm TCF}
-
J^{a,\downarrow}_{\rm TCF},
\label{eq:spin_channels}
\end{align}
where the overall normalization of the spin current is absorbed into its definition. All collinear altermagnets share a common symmetry operation, $[\overline{C}_2\Vert\mathcal{T}]$. This symmetry maps $\mathbf{k}$ to $-\mathbf{k}$ and transforms charge and spin numerators into their complex conjugates, ensuring that only the real part of $\mathcal N_{\kappa}(\mathbf k)$ that contributes to the total response. Additional spin-group symmetries are therefore required to enforce the vanishing of the charge response while allowing a finite spin response.\\
\indent A purely spatial mirror acts identically on the two conserved spin sectors and therefore imposes the same spatial parity on their sum and difference. It cannot by itself distinguish charge and spin-\(z\) currents. A complementary selection rule arises instead from a spin-group operation that combines the mirror with spin reversal. For \([C_2\Vert\mathcal M_x]\), where \(C_2\) exchanges the two conserved spin sectors,
\begin{align}
\mathcal N_0(\vk)
&\rightarrow
(-1)^{N_x}
\mathcal N_0(\mathcal M_x\vk),
\nonumber\\
\mathcal N_{s_z}(\vk)
&\rightarrow
-(-1)^{N_x}
\mathcal N_{s_z}(\mathcal M_x\vk).
\label{eq:spin_group_mirror}
\end{align}
Hence, for odd \(N_x\),
\begin{equation}
S^{a,0}=0,
\qquad
S^{a,s_z}
\ \text{allowed},
\label{eq:pure_spin_noSOC}
\end{equation}
which is the nonrelativistic counterpart of the mirror-selected pure-spin channel. By analyzing all 10 type-III collinear spin point groups that contain spatial inversion, we find that 9 of them host this symmetry operation (see SM for details). Therefore, most altermagnets with spatial inversion can generate spin currents based on a TCF. In the following, we present two representative examples to illustrate the above symmetry analysis.\\
\begin{figure}[!t]
    \centering
    \includegraphics[width=\columnwidth]{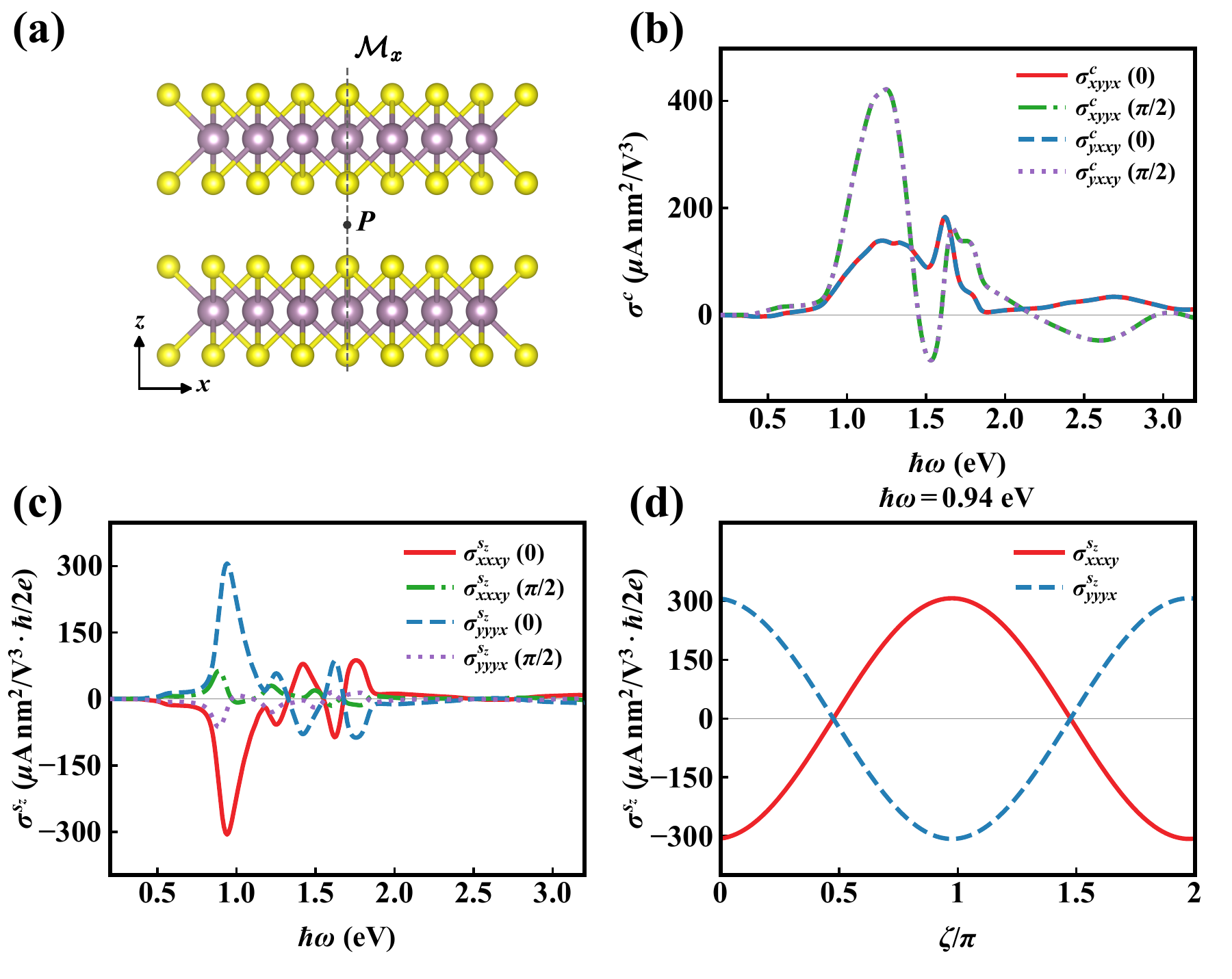}
    \caption{(a) Side view of the bilayer, showing the inversion
center $P$ and mirror plane $\mathcal{M}_x$.
(b),(c) Photon-energy dependence of the third-order dc
photoconductivities:
(b) charge components $\sigma_{xyyx}^{c}$ and $\sigma_{yxxy}^{c}$.
(c) $s_z$-polarized spin components $\sigma_{xxxy}^{s_z}$
and $\sigma_{yyyx}^{s_z}$.
Cartesian indices denote the current direction followed
by the field directions at $(\omega,\omega,-2\omega)$;
parentheses indicate the relative phase $\zeta$.
(d) Relative-phase dependence of $\sigma_{xxxy}^{s_z}$
and $\sigma_{yyyx}^{s_z}$ at
$\hbar\omega=0.94\,\mathrm{eV}$.
Charge and spin photoconductivities are expressed in
$\mu\mathrm{A}\mathrm{nm}^{2}/\mathrm{V}^{3}$ and
$\mu\mathrm{A}\mathrm{nm}^{2}/\mathrm{V}^{3}\cdot\hbar/(2e)$,
respectively.
}
    \label{fig:mos2_results}
\end{figure}
\indent\textit{Pure spin current in inversion-stacked bilayer MoS$_2$.---} We first apply the aforementioned symmetry criterion to inversion-stacked bilayer $2H$-MoS$_2$, a nonmagnetic system with strong SOC belonging to the gray point group $\bar{3}m1'=D_{3d}+\mathcal{T} D_{3d}$~\cite{Fang2015,Kormanyos2018}. As illustrated in Fig.~\ref{fig:mos2_results}(a), the bilayer preserves both inversion $\mathcal P$ and a vertical mirror $\mathcal M_x$. While inversion eliminates the conventional electric-dipole second-order responses, it leaves the third-order TCF response allowed. The additional mirror symmetry then provides the charge--spin selection rule derived above. For tensor components containing an odd number $N_x$ of $x$ indices among $(a,\alpha,\beta,\gamma)$, the charge contribution is odd under $\mathcal M_x$ and therefore vanishes after Brillouin-zone integration, whereas the corresponding $s_z$-polarized spin contribution is mirror even and remains symmetry allowed. Bilayer MoS$_2$ thus provides a centrosymmetric platform with nonnegligible SOC in which selected third-order TCF channels can carry a pure spin photocurrent.\\
\indent Figures~\ref{fig:mos2_results}(b) and \ref{fig:mos2_results}(c) show the nonzero third-order charge and spin TCF photoconductivities from a 12-band tight-binding model (see details in SM. According to the previous symmetry analysis, pure spin currents can be produced when the two polarization directions are parallel and perpendicular to this mirror plane, respectively. In this configuration, only four third-order tensor components, $xyyx$, $yyyx$, $yxxy$, and $xxxy$, remain nonzero. By varying the polarization configuration and the relative phase of the two-color field, we find that the spin current forms a pure-spin transport channel, while the accompanying charge current is allowed only along the perpendicular direction. The symmetry relations are also clearly reflected in the calculated responses: under the simultaneous interchange $x\leftrightarrow y$ of all Cartesian indices, the corresponding charge components retain the same sign, whereas the spin-current components acquire an additional minus sign. The first pronounced resonance occurs at a photon energy close to one half of the direct band gap at the high-symmetry $K$ point, consistent with the two-photon resonance condition of the $\omega$--$2\omega$ excitation. Furthermore, the spin current exhibits a periodic dependence on the relative phase of the two-color field. It is worth mentioning that the nonzero responses with a phase angle of $\pi/2$ all originate from finite relaxation times.\\
\indent\textit{Application to centrosymmetric altermagnets.---} We next consider a two-dimensional tetragonal two-sublattice model, adapted in simplified form from Ref.~\cite{Roig2024Minimal},
\begin{equation}
\begin{aligned}
H_{\rm AM}(\mathbf{k})={}&
t_1\cos\frac{k_x}{2}\cos\frac{k_y}{2}\tau_x\sigma_0
\\
&+t_2\sin k_x\sin k_y\tau_z\sigma_0
+J\tau_z\sigma_z .
\end{aligned}
\label{eq:altermagnetic_model}
\end{equation}
The model is inversion symmetric. In the absence of spin--orbit coupling, the two conserved spin blocks obey
\begin{equation}
\begin{aligned}
\tau_x h_s(k_x,k_y)\tau_x
&=h_{-s}(-k_y,k_x),\\
\tau_x h_s(-k_x,k_y)\tau_x
&=h_{-s}(k_x,k_y),
\end{aligned}
\label{eq:model_spin_group}
\end{equation}
corresponding to the spin-group operations $[C_2\Vert C_{4z}]$ and $[C_2\Vert\mathcal M_x]$, respectively. The former enforces the characteristic $d_{xy}$-type sign-changing altermagnetic spin splitting shown in Fig.~\ref{fig:altermagnet_results}(a), whereas the latter acts as a spin-flip mirror symmetry. Consequently, for tensor channels with odd $N_x$, the charge response is symmetry forbidden while the corresponding $s_z$ spin response remains allowed, yielding pure-spin tensor channels.\\
\begin{figure}[t]
    \centering
    \includegraphics[width=\columnwidth]{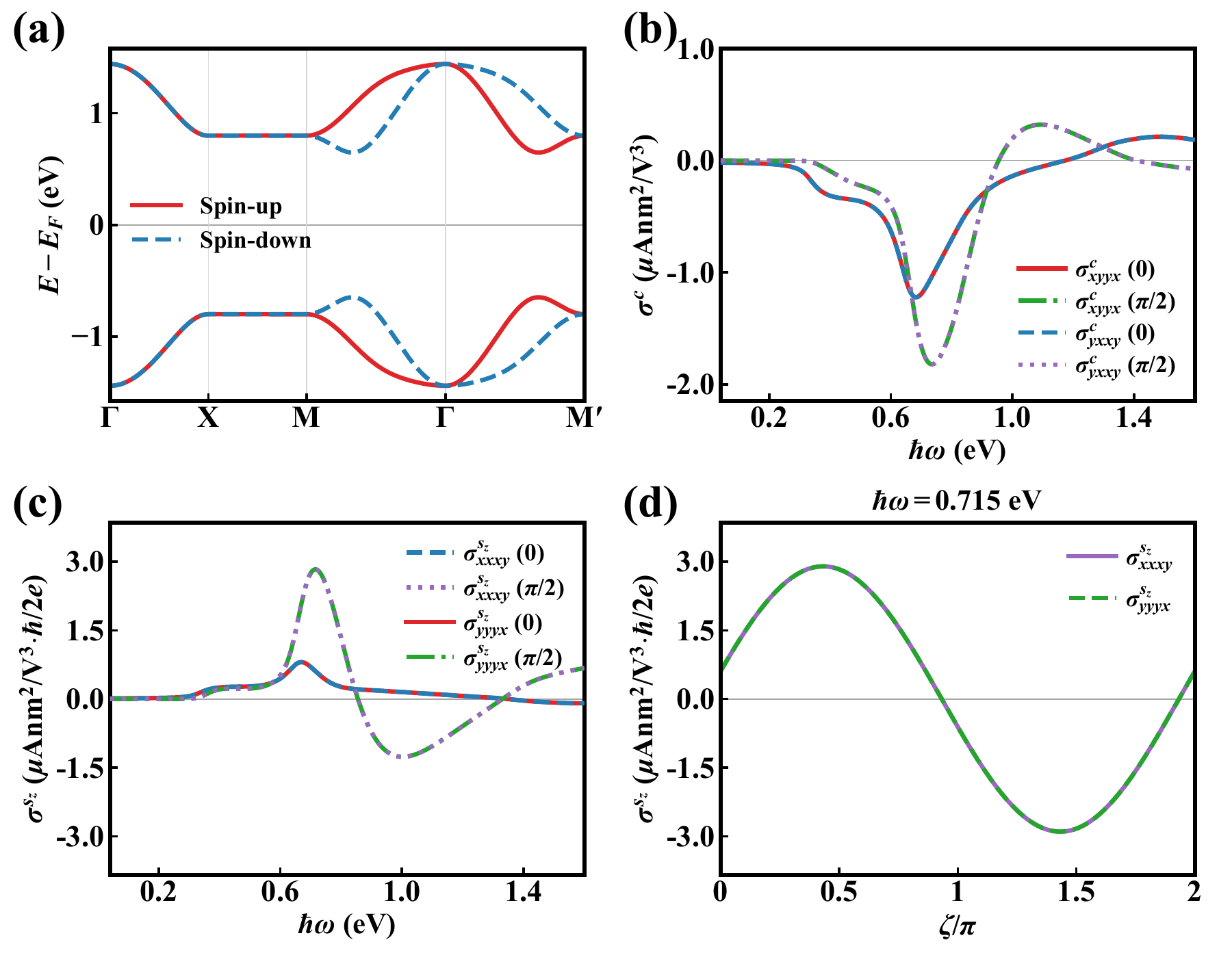}
    \caption{(a) Spin-resolved band structure of the centrosymmetric
    collinear altermagnet.
    (b) Photon-energy dependence of the symmetry-allowed
    third-order charge TCF photoconductivities
    $\sigma^{c}_{xyyx}$ and $\sigma^{c}_{yxxy}$.
    (c) Photon-energy dependence of the mirror-selected pure-spin
    photoconductivities
    $\sigma^{s_z}_{xxxy}$ and $\sigma^{s_z}_{yyyx}$.
    Angles in parentheses denote the relative TCF phase $\zeta$.
    (d) Relative-phase dependence of
    $\sigma^{s_z}_{xxxy}$ and $\sigma^{s_z}_{yyyx}$ at
    $\hbar\omega=0.715~\mathrm{eV}$.
    The equality of the symmetry-related Cartesian components at
    fixed $\zeta$ reflects the fourfold spin-group operation
    $[C_2\Vert C_{4z}]$.}
\label{fig:altermagnet_results}
\end{figure}
\indent Figures~\ref{fig:altermagnet_results}(b) and \ref{fig:altermagnet_results}(c) show the photon-energy dependence of the symmetry-selected charge and spin responses, respectively. The charge components $\sigma^{c}_{xyyx}$ and $\sigma^{c}_{yxxy}$ contain an even number of $x$ indices, and are therefore allowed by the spin-flip mirror $[C_2\Vert\mathcal M_x]$. In contrast, $\sigma^{s_z}_{xxxy}$ and $\sigma^{s_z}_{yyyx}$ have $N_x=3$ and $1$, respectively, so that their corresponding charge components are mirror odd and vanish, whereas the $s_z$ spin components remain symmetry allowed. These channels therefore realize pure-spin transport along the selected current directions. The calculated responses exhibit pronounced resonant structures, with a strong spin response near $\hbar\omega=0.715~\mathrm{eV}$. The symmetry-related Cartesian components coincide within numerical accuracy, reflecting the fourfold spin-group operation $[C_2\Vert C_{4z}]$. For the tensor components considered here, the resulting spatial and spin signs combine to give
\begin{align}
\sigma^{c}_{yxxy}(\zeta)
&=
\sigma^{c}_{xyyx}(\zeta),
\nonumber\\
\sigma^{s_z}_{yyyx}(\zeta)
&=
\sigma^{s_z}_{xxxy}(\zeta).
\label{eq:AM_C4_relations}
\end{align}
These relations are reproduced not only in the photon-energy dependence in Figs.~\ref{fig:altermagnet_results}(b) and \ref{fig:altermagnet_results}(c), but also in the relative-phase dependence in Fig.~\ref{fig:altermagnet_results}(d), where the two symmetry-related spin-current channels follow identical periodic variations with $\zeta$. The relative optical phase therefore continuously controls the magnitude and sign of the pure-spin response, while the spin-flip mirror $[C_2\Vert\mathcal M_x]$ determines the charge--spin selection rule. The model considered here has the same relevant spin-point-group symmetry as V$_2$Se$_2$O. However, the corresponding symmetry axes are rotated relative to the Cartesian axes adopted above. Consequently, the symmetry-selected orthogonal two-color polarization axes should be chosen along the in-plane diagonal directions, $[110]$ and $[1\bar{1}0]$, rather than along $x$ and $y$.\\
\indent\textit{Discussion.---} Our symmetry analysis shows that an $\omega$--$2\omega$ two-color field provides a general route to produce pure spin photocurrents in centrosymmetric systems. Beyond the orthogonal-polarization configuration considered above, parallel or symmetry-selected finite-angle polarizations can activate additional spin-current tensor components, while the relative phase $\zeta$ provides continuous control over their magnitude and sign. This mechanism is distinct from the band-geometry-driven spin photocurrent proposed in Ref.~\cite{Dong2026}. Although both are third-order responses, the latter has the form $\sigma^{(3)}(0;\omega,-\omega,0)$ and requires optical excitation together with a static electric field, whereas the present effect is an all-optical coherent-control response satisfying $\omega+\omega-2\omega=0$. It therefore requires no dc bias and enables phase-programmable spin-current generation. The same framework applies to both strong-SOC centrosymmetric magnets, described by magnetic point groups, and no-SOC altermagnets, described by spin groups, providing a broadly applicable route to resonantly enhanced and ultrafast pure spin photocurrents.\\
\indent Experimentally, the effect could be tested using overlapping, phase-locked $\omega$ and $2\omega$ pulses with orthogonal linear polarizations, following established pure-spin-current injection experiments~\cite{Stevens2003} and recent two-color studies in MoS$_2$ and halide perovskites~\cite{He2024,Xi2026}. Spin transport could be probed through Kerr microscopy of boundary spin accumulation~\cite{Kato2004,Anadon2025,Vavassori2026}, or inverse spin Hall conversion after injection into an appropriately oriented lateral detector~\cite{Kimura2007,Chen2024}. Quantitative interpretation requires accounting for spin relaxation, boundary conditions, and interface transmission. At fixed pulse envelopes, phase scans resolve the cosine and sine quadratures, while $J_s^{(3)}\propto I_\omega\sqrt{I_{2\omega}}$ provide complementary signatures in the weak-field, linear-readout regime. Phase cycling and single-beam controls help distinguish the coherent response from phase-independent backgrounds~\cite{Jayachandran2024}.\\
\begin{acknowledgments}
\indent This work is supported by the National Natural Science Foundation of China (Grant No. 12674150, and No. 12574070), the Postgraduate Scientific Research Innovation Project of Hunan Province (CX20250973).\\
\end{acknowledgments}
\bibliography{refs}

\end{document}